\documentclass[jap,superscriptaddress, preprintnumbers,twocolumn,showpacs]{revtex4}

\usepackage{graphicx}
\usepackage{dcolumn}
\usepackage{bm}
\usepackage{amsmath}
\usepackage{amssymb}
\usepackage{amsthm}
\usepackage{amsfonts}
\usepackage{algorithmic}
\usepackage{enumerate}
\usepackage{latexsym}
\usepackage{subfigure}
\usepackage{color}

\begin{document}

\preprint{AIP}

\title{Valley polarization in graphene-silicene-graphene heterojunction}

\author{Man Shen}
\affiliation{SKLSM, Institute of Semiconductors, Chinese Academy of
Sciences, P. O. Box 912, Beijing 100083, China}\affiliation{College of Physical Science and Information Engineering, Hebei Normal University, Shijiazhuang, Heibei 050016, China}
\author{Yan-Yang Zhang}
\affiliation{SKLSM, Institute of Semiconductors, Chinese Academy of Sciences, P. O. Box 912, Beijing 100083, China}
\author{Xing-Tao An}
\affiliation{School of Sciences, Hebei University of
Science and Technology, Shijiazhuang, Hebei 050018, China}\affiliation{Department of Physics and Center of Theoretical and Computational Physics, University of Hong Kong, Hong Kong, China}
\author{Jian-Jun Liu}
\affiliation{Physics Department, Shijiazhuang University, Shijiazhuang 050035, China}
\author{Shu-Shen Li}
\affiliation{SKLSM, Institute of Semiconductors, Chinese Academy of Sciences, P. O. Box 912, Beijing 100083, China}

\date{\today}

\begin{abstract}
Considering the difference of energy bands in graphene and silicene, we put forward a new model of the graphene-silicene-graphene (GSG) heterojunction. In the GSG, we study the valley polarization properties in a zigzag nanoribbon in the presence of an external electric field. We find the energy range associated with the bulk gap of silicene has a valley polarization more than $95\%$. Under the protection of the topological edge states of the silicene, the valley polarization remains even the small non-magnetic disorder is introduced. These results have certain practical significance in applications for future valley valve.
\end{abstract}

\maketitle

\section{\label{sec:level1}INTRODUCTION}

Graphene, the monolayer of carbon honeycomb lattice, has special electron and thermal transport properties\cite{K. S. NovoselovetNature,K. S. NovoselovetScience,A. K. Geim1,K. I. Bolotin,A. A. Balandin}. At the corners of the first Brillouin zone there are two degenerate and inequivalent valleys ($K$ and $K^{'}$). In momentum space two valleys have large interval, which leads to the strong suppression of the intervalley scattering\cite{A. F. Morpurgo,S. V. Morozovet,R. V. Gorbachevet}. Therefore the two valleys are proposed as independent internal degrees of freedom of the conduction electrons. The low-energy dynamics in the $K$ and $K^{'}$ valleys is given by the Dirac theory. In graphene the valley-dependent phenomena have attracted an increasing amount of interest\cite{D. Xiao,A. Rycerz1,J. M. Pereira,F. Zhai,D. Gunlycke}. The spin-orbit interaction in graphene is quite small, so the spin degeneracy can't be almost broken. Due to the smaller band gap in graphene, the good valley polarization only appears in a small energy range\cite{A. Rycerz,A. L. C. Pereira}. Therefore, it is hard to experimentally realize valleytronics in graphene.

Silicene, the monolayer of silicon, is isostructural to graphene\cite{A. K. Geim,A. H. Castro Neto} and has been experimentally synthesized\cite{P. Vogt,A. Fleurence}. Silicene has a strong spin-orbit interaction, and has a buckled sheet with two sublattices in two parallel planes.  These give rise to strong spin-valley dependence and valley Hall effect in silicene\cite{P. Vogt,C. C. Liu1,M. Ezawaprl,J. Y. Zhang}.
Applying an external electric field perpendicular to silicene's plane, the staggered potential between sublattices can be changed and the bulk gap can be tuned. In the bulk gap, there exist robust edge states connecting two valleys, giving rise to quantum spin Hall effect\cite{C. C. Liu1,M. Ezawaprl}. On the other hand, in the bulk band, the spin-valley configuration is quite different from that in graphene because of strong spin-orbital coupling\cite{Y. Y. Zhang}. It is therefore interesting to ask, what will the valley transport be like if graphene and silicene are connected together? Thereupon, we propose a graphene-silicene-graphene (GSG) heterojunction structure investigate the valley polarization through it.

In this paper, in the presence of an external perpendicular electric field we systematically investigate the properties of the valley polarization in graphene, silicene and GSG with zigzag edges, respectively, and the results are compared and analyzed. Under the four-band next-nearest-neighbor (NNN) tight-binding model, the Hamiltonian contains nearest neighbor (NN) hopping, the Rashba spin-orbit coupling term, the intrinsic spin-orbit coupling term and the staggered sublattice potential term. The Rashba spin-orbit coupling and staggered sublattice potential can both be tuned by the external electric field, which leads to the changes of the bulk band gaps and the spin split. Using the method of calculating transmission coefficient from an incident channel to an out-going channel and the recursion techniques\cite{T. Ando,T. Andoprb}, we obtain conductances in each valley. As we expect, the ideal valley polarization can appear within a larger energy range in the GSG. In addition, we find that in the GSG the valley polarization robust against the small non-magnetic disorder because of the protection of the topological edge states of the silicene.

This paper is organized as follows. The theoretical framework is introduced in Sec. \uppercase \expandafter {\romannumeral 2}. In Sec. \uppercase \expandafter {\romannumeral 3}, we present and discuss our results and then give a summary in Sec. \uppercase \expandafter {\romannumeral 4}.

\section{\label{sec:level1}THEORETICAL MODEL}
\begin{figure*}[!htbp]
\centering
\includegraphics[scale=2.2]{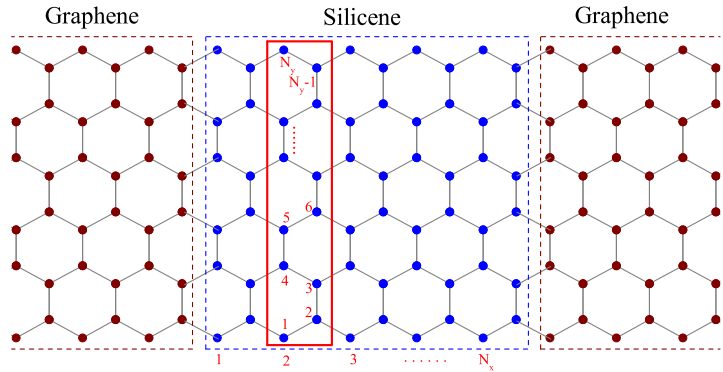}
\caption{\label{fig_1} Schematic illustration of GSG. The middle scattering region is the silicene sheet with the length $N_{x}$ and the width $N_{y}$, which is contacted by the graphene leads on the left and right. Honeycomb lattice of carbon or silicon atoms in a strip is with zigzag edges.}
\end{figure*}

The Hamiltonian of the silicene system can be described by the four-band NNN tight-binding model\cite{C. C. Liu,M. Ezawaprl,Y. Y. Zhang},
\begin{eqnarray}
H&=&t\displaystyle{\sum_{\langle{ij}\rangle,\alpha}}c_{i\alpha}^\dag c_{j\alpha}
-i\frac{2\lambda_{R}}{3}
\displaystyle{\sum_{\langle\langle{ij}\rangle\rangle,\alpha\beta}}\mu_{i}c_{i\alpha}^\dag
(\boldsymbol{\sigma}\times{\boldsymbol{\hat{d}}_{ij}})_{\alpha\beta}^{z}c_{j\beta}\nonumber\\
&&+i\frac{\lambda_{SO}}{3\sqrt{3}}\displaystyle{\sum_{\langle\langle{ij}\rangle\rangle,\alpha\beta}}
\nu_{ij}c_{i\alpha}^\dag\sigma_{{\alpha\beta}}^{z}c_{j\beta}+\lambda_{\nu}\displaystyle{\sum_{i,\alpha}}\xi_{i}c_{i\alpha}^\dag c_{i\alpha},\label{Eq1}
\end{eqnarray}
where $\langle{ij}\rangle$ and $\langle\langle{ij}\rangle\rangle$ denote all the NN and NNN hopping sites, respectively, and the indexes $\alpha,\beta$ label spin quantum numbers. The first term is the usual NN hopping with transfer energy $t = 1.6$eV for silicene, where $c_{i\alpha}^\dag$ creates an electron with spin polarization $\alpha$ at
site $i$. The second term describes the Rashba SOC between NNN sites, where $\mu_{i}=\pm1$ for the A(B) site and $\boldsymbol{\sigma}=(\sigma_{x},\sigma_{y},\sigma_{z})$ is the vector of the Pauli matrix of spin. $\boldsymbol{\hat{d}_{ij}}=\boldsymbol{d_{ij}}/|\boldsymbol{d_{ij}}|$ is the unit vector of $\boldsymbol{d_{ij}}$ which connects NNN sites $i$ and $j$. The third term represents the intrinsic SOC between NNN sites, where $\nu_{ij}=+1$ if the NNN hopping is anticlockwise with respect to the positive z axis and $\nu_{ij}=-1$ if it is clockwise. The fourth term describes the staggered sublattice potential term, and the parameter $\lambda_{\nu}=l_{z}E_{z}$ can be tuned by a perpendicular electric field $E_{z}$ because of the buckling distance $l_{z}$ between two sublattices. For silicene, the NN Rashba SOC can be ignored because it is very small and becomes zero at the gapless state\cite{M. Ezawaprl}. Thus, the main focus of this work is the NNN SOC terms and the staggered potential,
which can be tuned by the external electric field. For undoped graphene, the Hamiltonian is the first term of the Eq. (1) with $t = 2.7$eV, the very small intrinsic SOC and staggered potential term. Hereafter, we adopt the silicene's $t= 1.6$eV and lattice constant $a$ (NNN distance) as the units of energy and length, respectively.

The GSG is divided into three regions as
shown in Fig. 1, with left and right leads corresponding to graphene and the middle scattering region the silicene. Honeycomb lattices of carbon or silicon atoms in a strip are with zigzag edges, as shown in Fig. 1. In our numerical calculations, we fix the width of the conductor is 80 nanoribbons and each nanoribbon contains 80 atoms.

In Fig. 1 (and also in our calculations), the geometrical difference of lattice constants associated between graphene and silicene is ignored. The reasons are as follows. Firstly, in the calculation, this difference only reflects itself on the bond connecting configurations on the graphene-silicene interface. However, the well-accepted configurations and values of connecting bond hoppings at this interface from experiments or first principles calculations are lacking. On the other hand, even with the same geometric "lattice constant", sudden changes of Hamiltonian parameters across this interface is enough to induce strong scattering which we are interested in. Moreover, our calculations also offer intuitive pictures for cold atom systems where such structures with designed model parameters can be readily realized\cite{ColdAtom1,ColdAtom2}.

At low temperature, the conductance $G$ is given by the multichannel version\cite{D. S. Fisher} of Landauer's formula:
\begin{eqnarray}
G=\frac{2e^{2}}{h}\displaystyle{\sum_{\mu\nu}{|t_{\mu\nu}|}^2},\label{Eq2}
\end{eqnarray}
where $t_{\mu\nu}$ is the transmission coefficient from the incident channel $\nu$ with velocity $v_{\nu}$ to the out-going channel $\mu$ with velocity $v_{\mu}$, and can be calculated using the Green¡¯s function method in quasi-one-dimensional lattice\cite{T. Ando,Khomyakov2005}. The recursion techniques were employed in computing the Green functions\cite{T. Ando,T. Andoprb}.

The valley polarization of the transmitted current in $K$ valley and $K^{'}$ valley is defined by
\begin{eqnarray}
P_{KK^{'}}=\frac{G_{K}-G_{K^{'}}}{G_{K}+G_{K^{'}}},\label{Eq3}
\end{eqnarray}
and the polarization between difference valley is quantified by
\begin{eqnarray}
P_{intrainter}=\frac{G_{intra}-G_{inter}}{G_{intra}+G_{inter}},\label{Eq4}
\end{eqnarray}
where $G_{{K}(K^{'})}$ and $G_{intra(inter)}$ are the conductances transmitted to $K (K^{'})$ valley and between two same (difference) valleys, respectively.

In this paper, the parameters $\lambda_{R}=0t$, $\lambda_{SO}=0.01t$, $\lambda_{\nu}=0.001t$ are adopted for garaphene, and $\lambda_{R}=0.5t$, $\lambda_{SO}=0.5t$, $\lambda_{\nu}=0.05t$ for silicene. These spin-orbital parameters for silicene are rather larger than those in the realistic material, but this does not change the basic physics we will discuss\cite{Y. Y. Zhang}. As a matter of fact, we adopt them to manifest the physical consequences in our finite-size simulations.

\section{\label{sec:level1}RESULTS AND DISCUSSION}

\begin{figure*}[!htbp]
\centering
\includegraphics[scale=0.6]{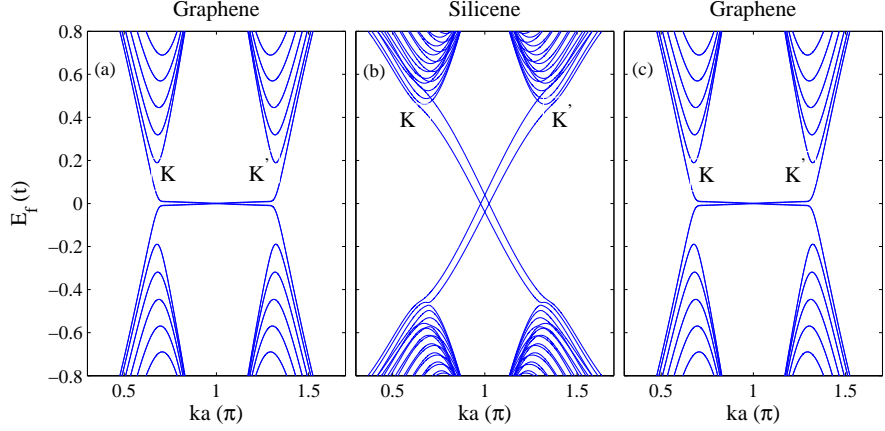}
\caption{\label{fig_2} Energy bands in a zigzag nanoribbon for graphene ((a) and  (c)) and for silicene ((b)).}
\end{figure*}

In Fig. 2 we show the energy bands obtained from diagonalizing the tight-binding Hamiltonian (1) with various parameters for a zigzag nanoribbon. When the Hamiltonian (1) has only the NN hopping energy with $t = 2.7$eV and vanishing Rashba SOC and staggered potential, the electronic structure exhibits a semimetallic behavior of the graphene, as shown in Fig. 2(a). In zigzag edge graphene, there are edge states connecting two valleys living in the lowest subband around Dirac points, whose energy extension is inversely proportional to the transverse width of the ribbon. Hence the transmission channels exist only in the $K^{'}$ valley in the vicinity of $E_f=0$ if the current is set to be right-going in the device. In silicene, with the finite Rashba SOC, intrinsic SOC and the staggered sublattice potential, gapless edge states appear in the bulk gap (see Fig. 2(b)), whose magnitude is determined by the staggered potential $2\lambda_{\nu}$. Experimentally, $2\lambda_{\nu}$ is tunable by a perpendicular electric field due to the buckled structure of two sublattices, therefore the bulk gap can be much bigger in silicene than the edge state region in graphene.

\begin{figure}[!htbp]
\centering
\includegraphics[scale=0.55]{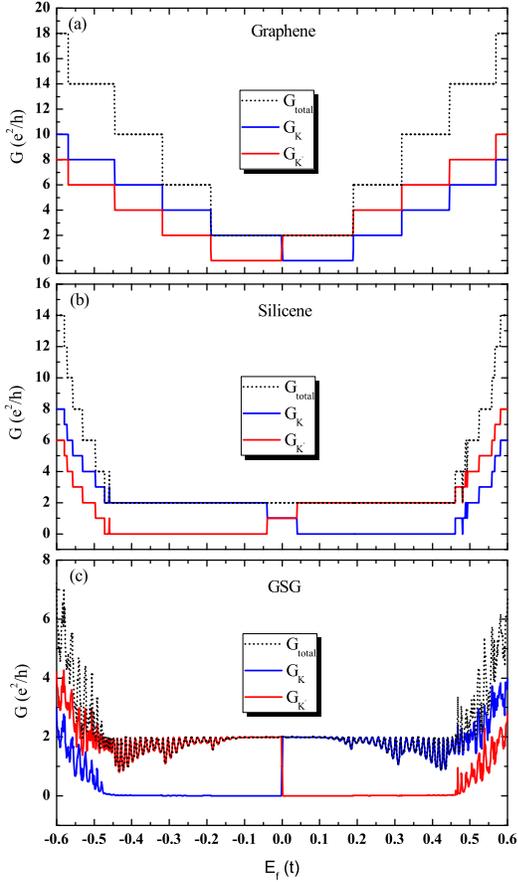}
\caption{\label{fig_3}The conductance as a function of the incident electron energy for zigzag edge geometries in graphene ((a)), silicene ((b)) and GSG ((c)). The red lines and the blue lines represent the conductance through $K$ valley and $K^{'}$ valley, respectively. The total conductance is indicated in black.}
\end{figure}

In Fig. 3, we show the normalized electrical conductance as a function of the incident electron energy for zigzag edge geometries in graphene ((a)), silicene ((b)) and GSG ((c)), respectively. For graphene (see Figs. 3(a)), the total conductance shows perfect quantized step-like plateaus and always increases by $4e^2/h$ since two bands start to transmit at the same time and the spin is degenerate. Consequently there are only conductance plateaus when $G/(2e^2/h)$ is even. Near zero energy, the electron transmits completely through the $K^{'}$ valley with $E_{f}>0$ and through the $K$ valley with $E_{f}<0$, which can be accounted for via the directions of the electronic velocity in the lowest energy subbands near $E=0$ (see Figs. 2(a)). Thereby the valley polarization can be produced in graphene\cite{A. Rycerz}. For the silicene with $\lambda_{R}=\lambda_{SO}=0.5$, $\lambda_{\nu}=0.05$, from the Fig. 3(b) we can see that the curves of the conductance still have obvious plateaus as depicted in graphene, with conductance plateau of $(2e^2/h)$ due to the topological edge states.

In GSG heterojunctions, the property of charge conductance changes much compared with those in graphene and silicene (see Fig. 3(c)). The obvious conductance plateaus disappear, which is caused by the mismatching of the interface between the graphene and the silicene. This lead to the oscillations of the total conductance and sharp dips at the edge of the conductance plateaus of the graphene that arise from the quantum interference between different spin channels in the GSG.
But the most interesting phenomenon is the almost perfect valley blockade in the bulk gap region ($-0.45t<E_f<0.45t$) of silicene: The electronic states in $K$ ($K^{'}$)) valley cannot transport at negative (positive) Fermi energy. This originates from the fact that at a definite Fermi energy in the bulk gap of silicene, the edge states around each valley possess identical velocity. Notice this blockade is effective even in the bulk band of graphene. Practically the bulk gap of silicene is more tunable and can be much larger than the level spacing of graphene from finite width, therefore this type of valley filter has a large and tunable working energy range compared with that from graphene itself\cite{A. Rycerz1}.

\begin{figure}[!htbp]
\centering
\includegraphics[scale=0.5]{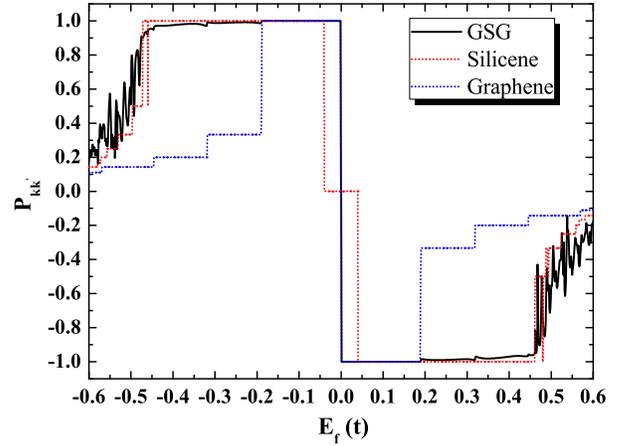}
\caption{\label{fig_4} The valley polarization of the transmitted current in $K$ valley and $K'$ valley as a function of the incident electron energy for graphene (blue line), silicene (red line) and GSG (black line).}
\end{figure}
For a more visualized view on the valley polarization, we plot the valley polarization of the transmitted current in $K$ valley and $K^{'}$ valley as a function of the incident electron energy in Fig. 4. For a pure graphene device, the plateaus of the full valley polarization is from $E_f=-0.1885t$ to $0.1885t$, just within the lowest subbands corresponding to the zigzag edge states. In other energy area the valley polarization decreases significantly. For contrast, the valleys of the edge states in silicene is also defined as those around Dirac point $K$ or $K^{'}$. Around $E_f=0$ there was no valley polarization for the pure silicene device (see the red line in Fig. 4). This can be attributed to the crossing of edge states with opposite velocities and spins, and the existence of spin-flip processes arising from nonzero $\lambda_{R}$\cite{XTAn}. However, in the GSG heterojunction, the valley polarization becomes more perfect than those in pure graphene or silicene devices. Thus, this GSG heterojunction is a good candidate for controlling the valley degree of freedom.

\begin{figure}[!htbp]
\centering
\includegraphics[scale=0.5]{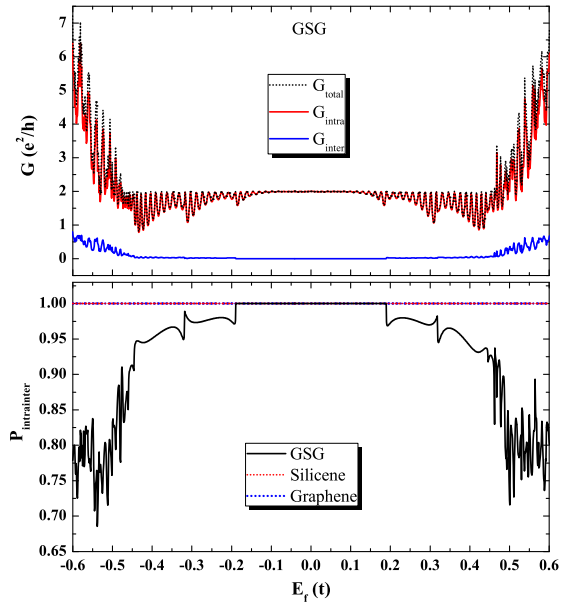}
\caption{\label{fig_5} (a) The conductance as a function of the incident electron energy between the same valley (red line), between the difference valley (blue line) and for totality (black dot line) in GSG. (b) The valley polarization of the transmitted current between difference valley as a function of the incident electron energy for graphene (blue dot line), silicene (red dot line) and GSG (black line).}
\end{figure}

In order to further investigate the stability of the valley polarization, in Fig. 5, the conductance and the valley polarization between two valleys as a function of the incident electron energy are plotted. Naturally, the inter-valley scattering in pure and clean graphene or silicene is vanishing. Nevertheless, in GSG, when the incident electrons lie within the bulk states of the graphene, they may be transmitted from one valley to another (see Fig. 5(a)), due to the strong scattering at the mismatched interface. But this kind of transmissive probability is very small, which makes the valley polarization between two difference valleys exceed $90\%$ from $E_f=-0.459t$ to $0.4606t$. So the valley polarization in $K$ valley and $K'$ valley can be better guaranteed.

\begin{figure}[!htbp]
\centering
\includegraphics[scale=0.4]{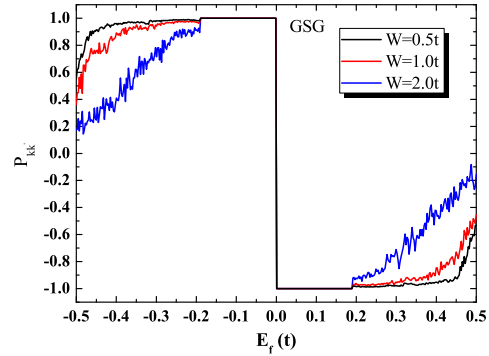}
\caption{\label{fig_6} The valley polarization $P_{KK^{'}}$ in GSG as a function of the incident electron energy for W=0.5t (black line), 1.0t (red line) and 2.0t (blue line). The results are the average over 100 disorder samples.}
\end{figure}

Additionally there are always other disorder effects in real material, e.g., impurities and defects. We investigate the non-magnetic disorder effect on the valley polarization in $K$ valley and $K'$ valley. Disordered on-site potential $W_i$ is added to each site $i$ in the central region, where $W_i$ is a random number uniformly distributed in the range $[-W/2, W/2]$ with the disorder strength $W$. Fig. 6 shows the valley polarization $P_{KK^{'}}$ versus the incident electron energy at various disorder strength for the GSG. From the black curve ($W=0.5t$) in Fig. (6), we can see that the valley polarization remains more than $90\%$ from $E_f=-0.45t$ to $0.45t$, which is due to the topological origin of the edge states. With the increasing of the disorder strength, the transmission of the carrier gradually becomes more weak and more chaotic, so the valley polarization becomes poor.

In summary, we proposed the GSG model, in which the conductance and the valley polarization are calculated. Using the tight-binding Hamiltonian, the energy bands for the graphene possess the spin degeneracy and a smaller bulk gap. On the contrary, in the silicene the spin degeneracy is lifted and the bulk gap increases. In the GSG heterojunction the valley polarization is very strong and corresponding to a wide energy range. In the GSG the carriers transmit mainly in the same valley, which ensures the stability of the valley polarization. The dependence of valley polarization on non-magnetic disorder is also discussed. These can make the GSG system be a good valley filter.

This work was supported by National Natural Science Foundation of China (Grant Nos. ??, ?? and ??).

\end{document}